\documentclass[runningheads]{llncs}
\usepackage[T1]{fontenc}
\usepackage{graphicx}
\usepackage{amssymb}
\usepackage{amsmath}
\usepackage{bm}
\usepackage{bbm}
\newcommand{\one}[1]{\mathbbm{1}_{[#1]}}
\usepackage{subcaption}
\usepackage{booktabs}
\usepackage[percent]{overpic}

\DeclareMathOperator{\MAP}{MAP}
\DeclareMathOperator{\Prec}{Prec}

\begin{document}
\title{
Contrastive Learning for Correlating Network Incidents
\thanks{Partially funded by the German Federal Ministry of Research, Technology and Space (BMFTR) in the FRONT-RUNNER project (Grant 16KISR005K).}
}
\author{Jeremias Dötterl
\orcidID{0009-0007-7801-4639}}

\authorrunning{J. Dötterl}
%
\institute{Adtran, Hermann-Dorner-Allee 91, 12489 Berlin\\
\email{jeremias.doetterl@adtran.com}}
\maketitle

\begin{abstract}
Internet service providers monitor their networks to detect, triage, and remediate service impairments.
When an incident is detected, it is important to determine whether similar incidents have occurred in the past or are happening concurrently elsewhere in the network.
Manual correlation of such incidents is infeasible due to the scale of the networks under observation, making automated correlation a necessity.
This paper presents a self-supervised learning method for similarity-based correlation of network situations.
Using this method, a deep neural network is trained on a large unlabeled dataset of network situations using contrastive learning.
High precision achieved in experiments on real-world network monitoring data suggests that contrastive learning is a promising approach to network incident correlation.

\end{abstract}

\section{Introduction}

Internet service providers monitor their networks for outages, performance degradations, and anomalies to maintain high service quality for their customers.
They use dedicated network management software to support monitoring, inspection, and troubleshooting of network devices.
This software aggregates monitoring data from numerous networking devices into a central location, such as a cloud data lake, and makes the data searchable, which allows the service provider to find devices and data of interest.

When an incident occurs, operators often investigate its scope by identifying other devices or time periods that exhibit similar behavior.
Moreover, if similar incidents have occurred in the past, correlating (matching) the new incident with the older incidents can accelerate finding root causes and choosing appropriate remedies.
While threshold-based queries are common in current practice, identifying structurally similar patterns (such as drops in a metric, repeating anomalies, or behavioral shifts) remains difficult and time-consuming, especially at scale.
In such environments, the operator is responsible for specifying the patterns in the monitoring data that serve as the correlation criterion.

This paper presents a method for correlating similar network situations based on \emph{learned similarity}.
The approach leverages recent advances in contrastive learning to embed network situations into a vector space where structurally related patterns are close together.
This enables operators to retrieve relevant incidents using an example-based query, rather than needing to define incident templates or similarity metrics manually.

\begin{figure}
\centerline{\includegraphics[width=0.78\linewidth]{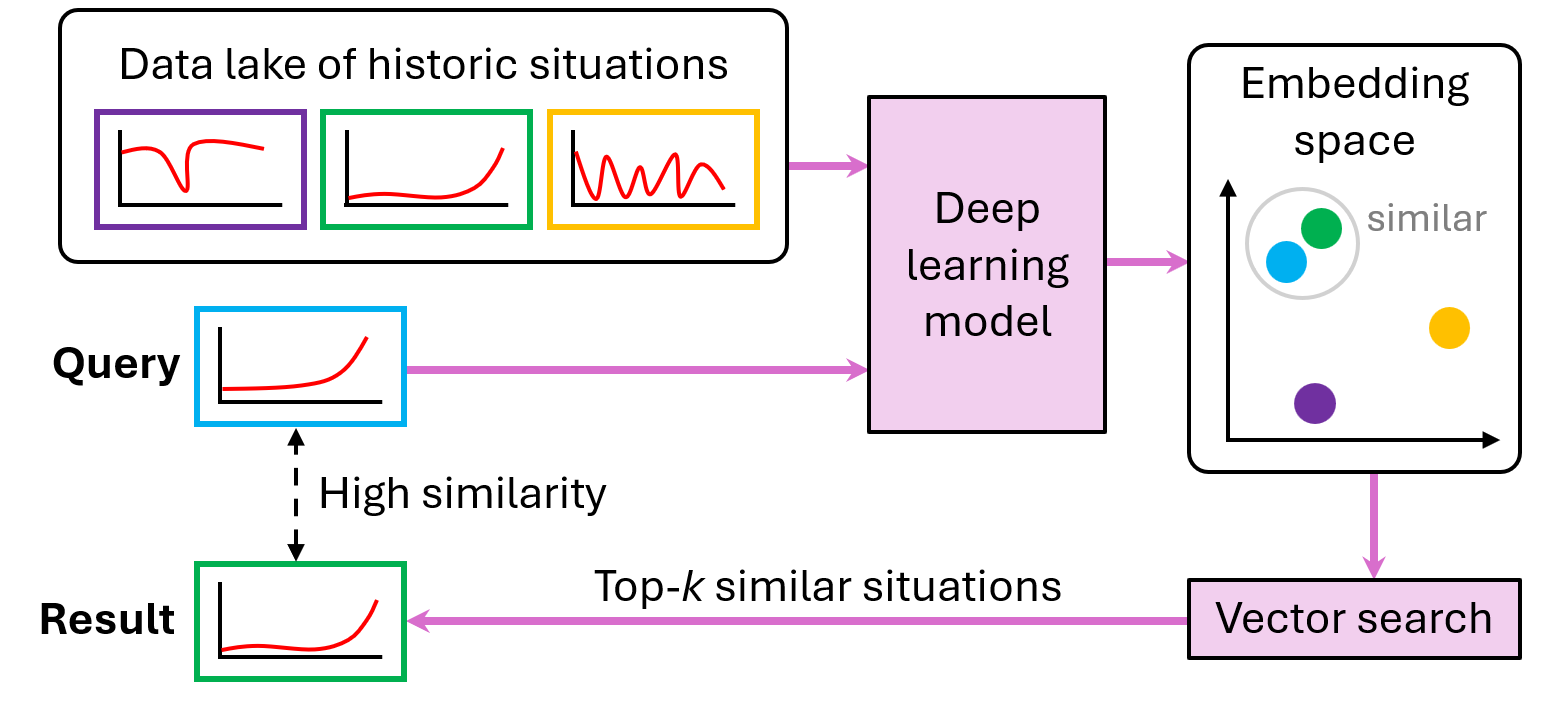}}
\caption{Contrastive Learning for Correlating Network Incidents}
\label{fig:method}
\end{figure}

The main idea is shown in Fig.~\ref{fig:method}.
A deep learning model is trained on a large set of network situations, stored in a cloud data lake.
The deep learning model is trained with a contrastive loss function such that similar situations have vector representations close to each other in the embedding space.
Given a reference network situation, the \emph{query}, the deep learning model is invoked to obtain an embedding for this query.
The top-\emph{k} most similar instances are retrieved via vector search, minimizing the cosine similarity between the vector embeddings.

This paper makes the following contributions.
First, it formalizes network incident correlation as a top-\emph{k} retrieval problem that can be approached with contrastive learning.
Second, it presents a solution method named \emph{Contrastive Learning for Situation Retrieval} (CLSR), which consists of a data preparation sequence and a deep learning model that is trained on a large set of unlabeled network situations.
Third, it evaluates CLSR on real-world networking telemetry from WLAN access points and shows that CLSR achieves high retrieval precision across diverse data patterns, suggesting that contrastive learning can serve as a promising paradigm for reliable network incident correlation.

The remainder of this paper is structured as follows.
Section \ref{sec:relatedwork} discusses related work in the areas of network incident correlation, alarm correlation, and contrastive learning.
The problem of correlating network incidents is formalized as a top-\emph{k} retrieval problem in section~\ref{sec:problem}.
Next, in section \ref{sec:method}, the contrastive learning method for network situation retrieval CLSR is presented.
Section \ref{sec:evaluation} contains a detailed evaluation of CLSR on a real networking dataset.
Finally, section \ref{sec:conclusion} draws conclusions and describes possible future work.

\section{Related Work}
\label{sec:relatedwork}

Correlating network incidents is an important task in network operations.
Traditional approaches often rely on alarm correlation, where systems analyze patterns in discrete alarms or alerts to infer related incidents \cite{julisch2001,julisch2003,salah2013}.
These methods include similarity-based rules, hierarchical rule systems, and statistical techniques \cite{mirheidari2013}.
Machine learning has also been applied, often through clustering or anomaly detection over alarm sequences.
In contrast, this paper focuses on learning incident similarity from raw monitoring data without requiring pre-defined event structures or alarm semantics.
This enables correlation across a broader range of patterns, including subtle data variations that may not trigger explicit alerts.

Recent works on incident correlation include Ban et al.~\cite{ban2023}, who propose an AI-assisted SIEM system that applies machine learning to reduce alert fatigue in security operations, and Freitas et al.~\cite{freitas2024}, who present GraphWeaver, a graph-based incident correlation engine for large scale deployments.
In cloud service environments, incident correlation has been approached using supervised and semi-supervised learning.
Chen et al.~\cite{chen2020identifying} propose LiDAR (Linked Incident identification with DAta-driven Representation), a deep learning method that identifies linked incidents using textual and structural signals.
Chen et al.~\cite{chen2022} propose GRLIA (Graph Representation Learning-based Incident Aggregation), a graph-based framework that uses unsupervised representation learning over monitoring-derived failure graphs.

The problem of learning similarity metrics for time series has been tackled with deep learning models such as NeuralWarp~\cite{grabocka2018}, which aligns time series through learnable warping functions.
Recent advances in contrastive learning, for instance, SimCLR~\cite{chen2020}, showed that strong embeddings can be learned through data augmentation.
Section~\ref{sec:method} describes a contrastive learning framework inspired by \cite{chen2020}, tailored to time series data in the context of network incident correlation.

\section{Top-\emph{k} Network Situation Retrieval}
\label{sec:problem}

This section formalizes network incident correlation as a retrieval problem.
For the scope of this paper, network situations are represented as time series, one of the most common and practical data types in network monitoring; the incorporation of networking topology and configuration is left to future work.

The problem of top-$k$ network situation retrieval is formalized as follows.
Let a network situation be a matrix in $\bbbr^{T \times C}$ that contains the values of $C$ features collected over $T$ time steps.
Given a large database $\mathcal{D}$ of network situations and a reference network situation $\bm X$ (the \emph{query}), the goal is to retrieve the top-$k$ situations in $\mathcal{D}$ that are \emph{most similar} to $\bm X$.
Similarity is measured by a function $\mathrm{Sim}(\bm X, \bm X'; \theta)$, where $\bm X$ and $\bm X'$ are network situations and $\theta$ are the parameters of a deep learning model. The function returns values from the unit interval.

The goal is to find values for $\theta$ such that $\mathrm{Sim}$ returns values near $1$ for similar network situation pairs and values near $0$ for dissimilar network situation pairs.
This similarity relationship is to be learned from a large set of example pairs.

\section{Contrastive Learning for Situation Retrieval (CLSR)}
\label{sec:method}

This section introduces \emph{Contrastive Learning for Situation Retrieval} (CLSR). 

Let $M_\theta$ denote a deep learning model $M$ with parameters $\theta$ that computes a vector embedding of $\bm X$.
Let $\mathrm{cos\_sim}(\bm a, \bm b)$ denote the cosine similarity of two vectors $\bm a, \bm b$ defined as $\frac{\bm a \cdot \bm b}{\|\bm a\| \, \|\bm b\|}$.
We choose $\mathrm{Sim}(\bm X, \bm X' ; \theta)$ to be the cosine similarity between the embeddings of the network situations $(\bm X, \bm X')$: 
\[
\mathrm{Sim}(\bm X, \bm X'; \theta) = \mathrm{cos\_sim}(M_\theta(\bm X), M_\theta(\bm X'))
\]

This choice allows us to approach the retrieval of similar network situations with contrastive learning:
\begin{enumerate}
    \item Prepare a data set of similar network situation pairs
    \item Train the deep learning model $M_\theta$ using a contrastive loss function
\end{enumerate}

\subsection{Data preparation}
\label{sec:datapreparation}
A dataset has to be prepared that contains a sufficient number of similar network situation pairs $\{(\bm X, \bm X')\}$.
As the proposed method learns similarity relationships from this dataset, the content of this dataset determines which network situations will be considered similar by the model.
In our setting, we focus on multivariate time series of $T$ time steps and $C$ features.

The following transformations can be applied to the data to obtain a suitable contrastive learning data set.

\subsubsection{Odd-even splitting}

Consider a sequence of $2T$ feature vectors:
\[
\mathbf{s}_{1},\, \mathbf{s}_{2},\, \ldots,\, \mathbf{s}_{2T}, \quad \mathbf{s}_{i} \in \bbbr^C.
\]

A pair of similar network situations \( \mathbf{X},\, \mathbf{X}' \in \bbbr^{T \times C} \) can be obtained by stacking the odd-indexed vectors and the even-indexed vectors into two matrices:
\[
\mathbf{X} =
\begin{bmatrix}
    \mathbf{s}_{1}^\top \\
    \mathbf{s}_{3}^\top \\
    \vdots \\
    \mathbf{s}_{2T-1}^\top
\end{bmatrix}
\qquad
\mathbf{X}' =
\begin{bmatrix}
    \mathbf{s}_{2}^\top \\
    \mathbf{s}_{4}^\top \\
    \vdots \\
    \mathbf{s}_{2T}^\top
\end{bmatrix}
\]

This results in pairs that are structurally similar but not identical, which is necessary for the model to learn meaningful similarity relationships.

\subsubsection{Imputation}
In network monitoring, missing values are often indicators that a device is disconnected or powered off.
For missing values, a special value is imputed that is known to be outside the usual data range for that feature.

\subsubsection{Optional transformations}
The following transformations can optionally be applied to influence the type of similarities the model detects.
They are referred to as optional because, as shown in the evaluation experiments (Section~\ref{sec:evaluation}), they do not consistently improve retrieval precision across all queries.
Instead, they increase precision for some queries while reducing it for others.
Therefore, they can be used deliberately to steer similarity detection.

\paragraph{Random cyclic shift}

Shifts the time series along the time axis by a random amount, wrapping around at the boundaries.
This transformation is intended to make the model less sensitive to absolute temporal alignment, allowing it to recognize similarity based on shape regardless of the specific position in time.

\paragraph{Random vertical shift}
Adds a random offset to the time series.
This transformation is intended to make the model less sensitive to baseline shifts.

\paragraph{Random scaling}
Multiplies the time series by a randomly selected factor.
The goal is to reduce the model's reliance on absolute signal magnitudes and instead emphasize the relative shape and dynamics of the data.

\subsection{Model training and contrastive loss}

The deep learning model is trained using mini-batch stochastic gradient descent on batches of size $B$ with a contrastive loss function.
A deep learning architecture is chosen for $M_\theta$ that accepts tensors of shape $(B, T, C)$.
That is, the model takes a batch of $B$ network situations, each of shape $(T, C)$, and outputs a matrix $(B, E)$, which contains an $E$-dimensional vector embedding for each situation in the batch.
Many model architectures can be viable; one suitable architecture is explored in the evaluation in section~\ref{sec:evaluation}.

For the loss, we closely follow the work of Chen \emph{et al.} \cite{chen2020}.
We sample $B$ data pairs randomly.
For each pair $(i, j)$, we generate their embeddings $\bm z_i, \bm z_j$ and
calculate the normalized temperature-scaled cross entropy loss \cite{sohn2016,wu2018,oord2018,chen2020}:

\begin{equation}
\label{eq:loss}
    \ell_{i,j} =
    -\log
    \frac{
        \exp(\mathrm{cos\_sim}(\bm z_i, \bm z_j)/\tau)
    }{\sum_{k=1}^{B}
    \one{k \neq i}\exp(\mathrm{cos\_sim}(\bm z_i, \bm z_k)/\tau)}
\end{equation}
where $\one{p}$ is the indicator function, which yields $1$ if the predicate $p$ is true and $0$ otherwise; and $\tau$ is a temperature hyperparameter. 
We compute the loss for all positive pairs $(i, j)$ and $(j, i)$, and the mean of all losses is used as the final loss of the batch.

\section{Evaluation}
\label{sec:evaluation}
This section evaluates the performance of CLSR on a real networking data set.

\subsection{Metrics and baseline}
We evaluate retrieval performance over $n$ tasks.
Each task consists of a reference situation $\mathbf{X}_i$ and a set of ground-truth network situations $\mathcal{R}_i$ similar to $\mathbf{X}_i$.
The full dataset $\mathcal{D}$ includes all ground-truth situations $\bigcup_i \mathcal{R}_i$ and additional network situations that are not part of any task, serving as distractors during retrieval.

For task $i$, let $A_i = (A_{i,1}, \ldots, A_{i,k})$ be the $k$ retrieved items.
Performance is measured with \emph{Mean Average Precision} (MAP) \cite{manning2008}:

\[
\MAP(k)
= \frac{1}{n}
  \sum_{i=1}^{n}
    \left(
      \frac{1}{|\mathcal{R}_i|}
      \sum_{r=1}^{k}
        \Prec_i(r)\,
        \one{A_{i,r} \in \mathcal{R}_i}
    \right),
\]
where the precision at rank $r$ is
$
\Prec_i(r)
= \frac{1}{r}
  \sum_{j=1}^{r}
    \one{A_{i,j} \in \mathcal{R}_i}
$.

As a baseline, a retriever is used that returns the network situations with the smallest L2 distance from the reference situation, i.e., the retriever minimizes
$\|\operatorname{vec}(\mathbf{X}) - \operatorname{vec}(\mathbf{X'})\|_2$
where $\operatorname{vec}$ is an operator that flattens a matrix to a vector.
This baseline is simple to implement and does not require any training on data.
Therefore, we require CLSR to significantly outperform this baseline to justify the higher complexity of CLSR compared to this baseline.

\subsection{Data sets}

For this evaluation, a large data set of training and validation data is needed to train the deep learning model.
This data set can be unlabeled, which is a strength of our self-supervised learning approach.
Additionally, a small labeled data set is needed to measure the retrieval performance.

\subsubsection{Data source}
To evaluate CLSR under realistic conditions, a real networking data is used that was collected from WLAN access points (APs) over a period of 31 days.
These APs belong to hundreds of test installations on which data is captured for networking research; the data is stored in anonymized form with the users' consent.
The AP collects in 10-second intervals the retransmission rate for each of the associated client devices.
The retransmission rate is a value between 0 and 100, indicating the percentage of IEEE802.11 frames that were retransmitted during the 10-second interval.

\subsubsection{Data preprocessing}

We focus on retransmission patterns over 60-minute intervals.
Each retransmission time series is divided into 60-minute segments, and the 10 samples within each 60-second window are averaged to produce a 60-point time series.
Since devices can disassociate and reassociate with the AP at any time, some of these 60 values may be missing.

Time series from 2 of the 31 days were set aside to be used as test data, as described in the next subsection.  
To the time series from the remaining 29 days, the treatment described in Section~\ref{sec:datapreparation} was applied.  
The odd-even splitting procedure was performed to create pairs of similar time series.  
Through this process, two time series were obtained, each of length $T=30$.  
Time series with fewer than $10$~data points were discarded.  
Imputation was performed by replacing missing values with $-100$, a value outside the valid range for retransmissions.  
For a subset of the experiments, the optional transformations \emph{random cyclic shift}, \emph{vertical shift}, and \emph{scaling} were applied to the second time series of each pair to measure their effect on retrieval performance:
\begin{itemize}
    \item Vertical shift: The time series was shifted by an offset randomly drawn from the uniform distribution over the interval $(-10, 10)$.
    \item Scaling: The time series was scaled by a factor randomly drawn from the uniform distribution over the interval $(0.5, 2.0)$.
\end{itemize}
After this preprocessing, the data was split into a training set of 876130~samples and a validation set of 33678~samples, which are used to train the deep learning model and to guide hyperparameter search.

\subsubsection{Labeled test data}

\begin{figure}
\centering
\begin{overpic}[width=0.95\linewidth]{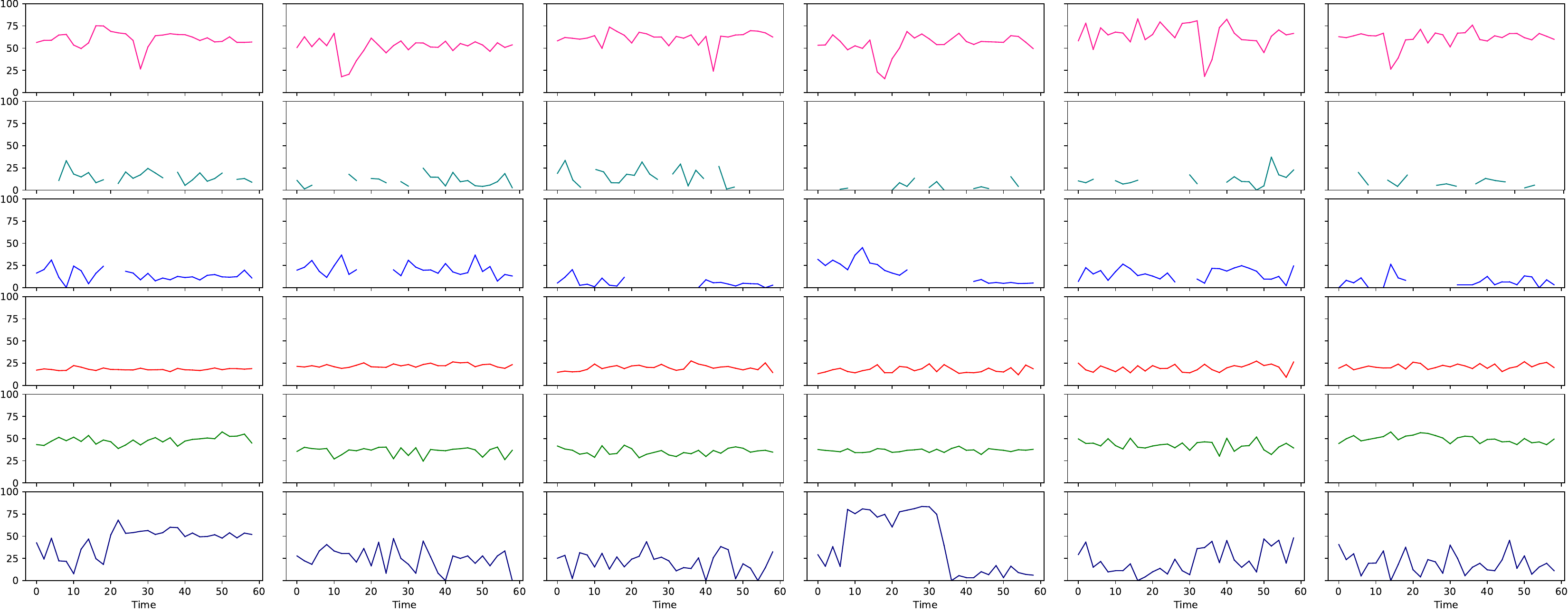}
    \put(50, -1.5){Time}
    \put(-2.5, 10){\rotatebox{90}{Retransmissions}}
\end{overpic}
\caption{Examples from the labeled test dataset, each row showing another class: 
Drop (row~1),
multiple disassociations (row~2),
singe disassociation (row~3),
stable around 20\% (row~4),
stable around 40\% (row~5),
and others (row~6).}
\label{fig:class-examples}
\end{figure}

To obtain a labeled test dataset, plots of the time series were manually inspected and time series were identified that matched any of the following classes; examples for each of the classes are shown in Fig.~\ref{fig:class-examples}:
\begin{itemize}
    \item \emph{Drop}: The retransmission rate dropped from a high level to a significantly lower level for a short period before it rises again to its initial level.
    \item \emph{Multiple disassociations}: The client disassociated multiple times, indicated by a gap in the time series.
    \item \emph{Single disassociation}: The client disassociated once for multiple minutes.
    \item \emph{Stable around 20\%}: The retransmission rate remains near 20~percent, with only minor fluctuations over time.
    \item \emph{Stable around 40\%}: The retransmission rate remains near 40~percent, with only minor fluctuations over time.
\end{itemize}
These classes correspond to notable patterns that could be visually identified by inspecting the time series.
Via this manual labeling approach, 40~labeled samples could be obtained, with 8~samples per class.
Labeling was time consuming, but fortunately is only needed for the evaluation, not for the training.

From these labeled samples, 40~retrieval tasks were derived as follows:
For each of the 8~members of each class, this member was selected as the query $\mathbf{X}$ and the remaining 7~elements as the corresponding ground truth set $\mathcal{R}$.

Additional 48~timeseries were selected from the data set that did not correspond to any of the classes.
They are shown in the last row of Fig.~\ref{fig:class-examples}.
They are used to enlarge the labeled dataset with irrelevant samples that are never supposed to be retrieved.

\subsection{Models and training}

For evaluation, the deep learning model architecture shown in Table \ref{tab:model} was used.
The model consists of a sequence of convolution layers with a kernel size of 5 and densely connected layers with ReLU activation.
After each convolution, batch normalization is applied.
A dropout \cite{srivastava2014} layer with a dropout rate of 50\% is used to prevent overfitting.
This model architecture was found by exploring different architectures in informal experiments.

\begin{table}[h!]
\setlength{\tabcolsep}{6pt}
\centering
\caption{Model architecture with batch size $B=256$ and sequence length $T=30$}
\begin{tabular}{ll}
\toprule
\textbf{Layer type} & \textbf{Output shape} \\
\midrule
InputLayer & (B, T, 1) \\
Normalization & (B, T, 1) \\
Conv1D($\text{kernel}=5, \text{stride}=1$) & (B, T, 128) \\
BatchNormalization & (B, T, 128) \\
ReLU & (B, T, 128) \\
Conv1D($\text{kernel}=5, \text{stride}=1$) & (B, T, 128) \\
BatchNormalization & (B, T, 128) \\
ReLU & (B, T, 128) \\
Conv1D($\text{kernel}=5, \text{stride}=1$) & (B, T, 128) \\
BatchNormalization & (B, T, 128) \\
ReLU & (B, T, 128) \\
Dense & (B, T, 128) \\
Dropout($\text{rate}=0.5$) & (B, T, 128) \\
GlobalAveragePooling1D & (B, 128) \\
Dense & (B, 128) \\
\bottomrule
\end{tabular}
\label{tab:model}
\end{table}

The evaluation compares eight models, which were trained on different data variants and with different values for temperature $\tau$.
The model names and configurations are shown in Table~\ref{tab:models}.

\begin{table}[h!]
\setlength{\tabcolsep}{6pt}
\centering
\caption{Model configurations}
\begin{tabular}{lcccc}
\toprule
\textbf{Model} & \textbf{$\tau$} & \textbf{Cyclic shift} & \textbf{Vertical shift} & \textbf{Scaling} \\
\midrule
CLSR-10 & $0.10$ & No & No & No \\
CLSR-20 & $0.20$ & No & No & No \\
CLSR-10-cyclic-shift & $0.10$ & Yes & No & No \\
CLSR-20-cyclic-shift & $0.20$ & Yes & No & No \\
CLSR-10-vertical-shift & $0.10$ & No & Yes & No \\
CLSR-20-vertical-shift & $0.20$ & No & Yes & No \\
CLSR-10-scale & $0.10$ & No & No & Yes \\
CLSR-20-scale & $0.20$ & No & No & Yes \\
\bottomrule
\end{tabular}
\label{tab:models}
\end{table}

All models were trained with a batch size of $B=256$ using the AdamW \cite{loshchilov2018} optimizer with a learning rate of $1e^{-5}$ and a weight decay of $1e^{-4}$.
The Early Stopping method was applied with a patience of 5~epochs to prevent overfitting, i.e., training was stopped once the validation loss did not improve for 5~consecutive epochs.
The model checkpoint with the lowest validation loss was used to measure the retrieval accuracy.

\subsection{Results}

Table~\ref{tab:precision-at-k} shows \emph{MAP} and \emph{Precision@k} for the different model variants and the baseline for different values of \emph{k}.
It can be observed that all variants of CLSR strongly outperform the L2Retriever baseline.
CLSR-10-scale achieves a MAP of $0.908$, whereas the baseline reaches $0.533$.
Looking at the precision values, the best performing variants of CLSR reach perfect precision for @1, while the L2Retriever only achieves a precision of $0.525$.

\begin{table}
\setlength{\tabcolsep}{6pt}
\centering
\caption{MAP and Precision@k (k = 1, 3, 5) for each retriever.}
\label{tab:precision-at-k}
\begin{tabular}{lcccc}
\toprule
\textbf{Retriever} & \textbf{MAP} & \textbf{@1} & \textbf{@3} & \textbf{@5} \\
\midrule
L2Retriever (baseline) & 0.533 & 0.525 & 0.600 & 0.530 \\
CLSR-10 & 0.883 & \textbf{1.000} & 0.908 & 0.860 \\
CLSR-20 & 0.903  & 0.975 & 0.933 & 0.880 \\
CLSR-10-cyclic-shift & 0.894 & \textbf{1.000} & 0.942 & 0.890 \\
CLSR-20-cyclic-shift & 0.895 & 0.975 & 0.933 & \textbf{0.895} \\
CLSR-10-vertical-shift & 0.899 & 0.975 & 0.908 & 0.870 \\
CLSR-20-vertical-shift & 0.877 & 0.950 & 0.892 & 0.865 \\
CLSR-10-scale & \textbf{0.908} & \textbf{1.000} & \textbf{0.967} & 0.890 \\
CLSR-20-scale & 0.867 & 0.975 & 0.900 & 0.850 \\
\bottomrule
\end{tabular}
\end{table}

\begin{figure}
\centerline{\includegraphics[width=0.98\linewidth]{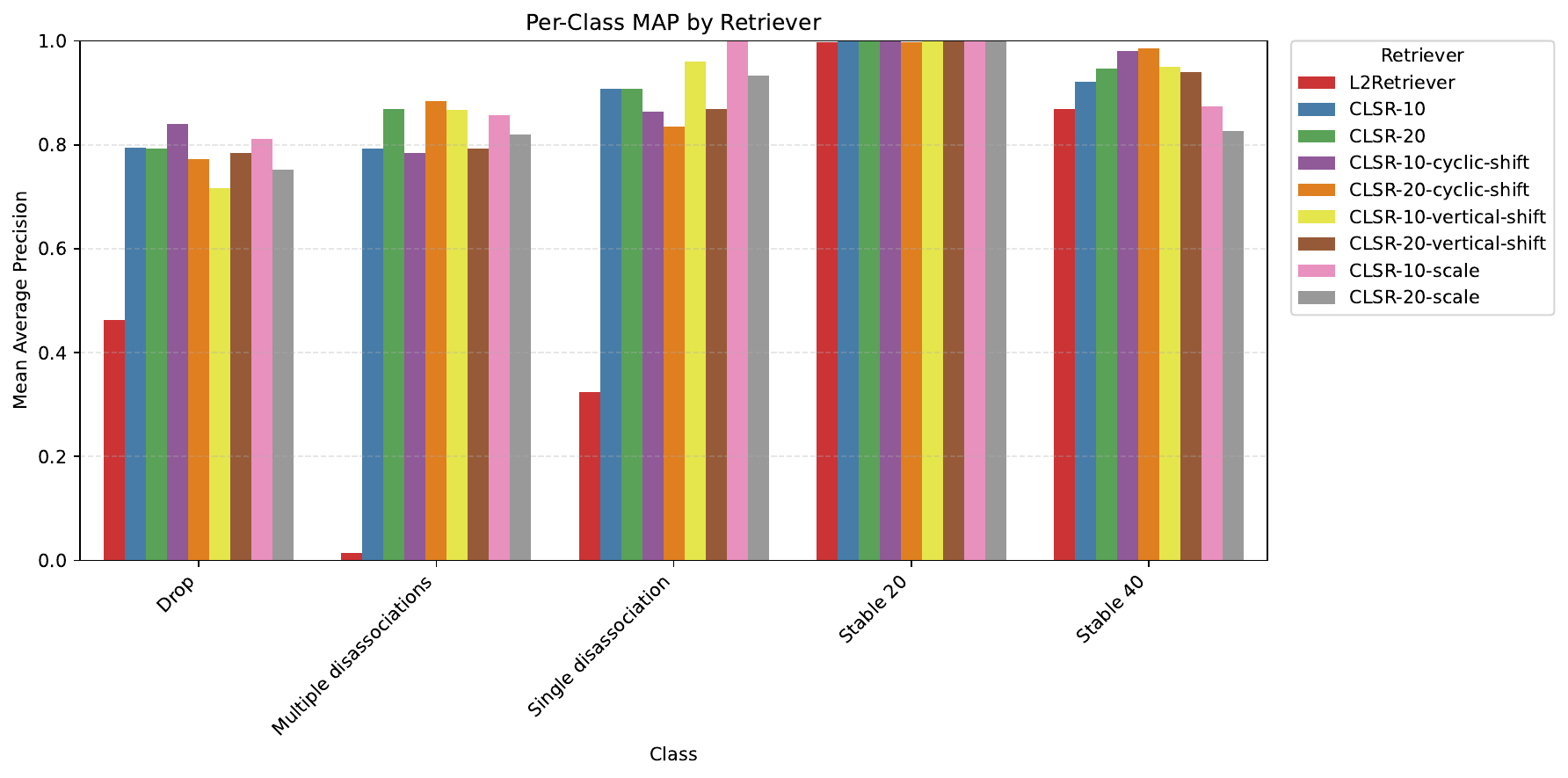}}
\caption{Per-class MAP by retriever}
\label{fig:results}
\end{figure}

The different CLSR variants achieve varying retrieval results, which can be better understood by examining the per-class mean average precision (MAP) for each retriever, as shown in Fig.~\ref{fig:results}.

\begin{itemize}
    \item
    The \emph{Stable 20} class is retrieved accurately by all retrievers, indicating it is relatively easy to detect.
    In contrast, the \emph{Stable 40} class poses a greater challenge, likely because retransmissions at this level are less consistent than those at 20\%.
    Furthermore, applying transformations such as scaling or vertical shifts appears to reduce retrieval performance for this class.
    \item
    The \emph{Multiple disassociations} class is particularly difficult for the L2 retriever.
    This class seems to require robustness to both horizontal and vertical shifts, which the L2 baseline lacks.
    CLSR variants, however, are better equipped to handle such variability.
    \item
    Training on augmented data can improve performance for certain classes.
    For instance, the CLSR-10-scale variant outperforms the vanilla CLSR on the \emph{Single disassociation} class.
    However, no single augmentation method consistently outperforms all others across the board.
    Instead, each augmentation encodes a different similarity preference, resulting in trade-offs in performance across classes.
    \item
    The temperature parameter $\tau$ also influences retrieval performance.
    While it affects how tightly or loosely similarity is enforced during training, there is no universally optimal value.
    A setting that improves performance on one class may reduce it on another, again indicating the presence of trade-offs.
\end{itemize}

Overall, CLSR consistently outperforms the L2Retriever baseline across most classes, though no single CLSR variant is best in every case.
It is important to note that all models were trained on a fully unlabeled dataset.
Further improvements in retrieval performance could potentially be achieved through supervised fine-tuning on labeled data, or by experimenting with additional model architectures and hyperparameters.

\section{Conclusions}
\label{sec:conclusion}

This paper presented CLSR, a contrastive learning approach for automated correlation of network incidents based on similarities in the monitoring data.
By training a deep neural network to produce embeddings that reflect similarity between network situations, CLSR enables effective similarity-based retrieval without requiring labeled data or hand-crafted features.

The evaluation on real-world networking data demonstrates that CLSR significantly outperforms a baseline L2 retriever, particularly for complex classes that require invariance to noise, temporal shifts, or scale.
While no single variant of CLSR is best across all classes, CLSR provides flexibility through data augmentation and temperature tuning to reflect different similarity preferences.
The evaluation suggests that learning situation embeddings with contrastive loss can be a viable path towards reliable network incident correlation.

Potential future work includes fine-tuning CLSR on labeled data and extending CLSR to additional data structures beyond time series.
For example, an integration with topology graphs or tabular configuration data would be a valuable extension.

\bibliographystyle{splncs04}
\bibliography{bib}

\end{document}